\documentclass[11pt]{article}

\usepackage[margin=1in]{geometry}
\usepackage{microtype}
\usepackage{setspace}
\usepackage{amsmath,amssymb}
\usepackage{graphicx}
\usepackage{booktabs}
\usepackage{tabularx}
\usepackage[numbers,sort&compress]{natbib}
\usepackage[hidelinks]{hyperref}
\usepackage{authblk}

\title{Semi-Empirical Pulsation Reconstruction of Delta Cephei with Photometry, Radial Velocities, and Temperature Constraints}

\author[1,2,*]{Zuhoor Elahi}
\author[1]{Wafa Gull}

\affil[1]{Department of Physics and Astronomy, University of Southern Mississippi, Hattiesburg, MS, USA}
\affil[2]{Department of Physics, University of Karachi, Karachi, Pakistan}
\affil[*]{Corresponding author: zuhoor.elahi@usm.edu}

\date{}

\begin{document}
\maketitle

\begin{abstract}
We present a semi-empirical reconstruction of the pulsation behavior of the classical Cepheid \(\delta\) Cephei using observed Johnson-\(V\) photometry, HARPS-N radial velocities, and published SPIPS pulsation-model curves. The goal is to construct an observationally anchored benchmark for separating the roles of light-curve morphology, radial displacement, effective-temperature variation, and absolute radius scale. A cleaned AAVSO Johnson-\(V\) data set was phase folded with \(P=5.366531\,{\rm d}\), corrected for observer zero-point offsets, clipped for residual outliers, and fitted with a three-harmonic Fourier template. The resulting empirical visual template has \(\Delta V \simeq 0.833\,{\rm mag}\), \(R_{21}\simeq 0.382\), \(R_{31}\simeq 0.168\), and a rise fraction of \(\simeq 0.287\). After phase and vertical alignment, the published SPIPS \(V\)-band curve reproduces the empirical Johnson-\(V\) morphology with an RMS residual of \(\simeq 0.023\,{\rm mag}\). A Fourier representation of the HARPS-N radial velocities gives a peak-to-peak velocity span of \(\simeq 40.82\,{\rm km\,s^{-1}}\), corresponding to a preliminary radius-displacement amplitude of \(\simeq 4.81\,R_\odot\) for \(p=1.317\). Combining this radius curve with the SPIPS \(T_{\rm eff}(\phi)\) curve yields a hybrid luminosity curve close to the published SPIPS luminosity curve. The adopted \(R_0=43.7\,R_\odot\) scale gives a mean luminosity ratio \(L_{\rm hybrid}/L_{\rm SPIPS}\simeq 1.04\), while \(R_0=44.9\,R_\odot\) gives \(\simeq 1.10\). A direct radius-scale comparison similarly favors the lower adopted radius scale, with an RMS offset of \(0.864\,R_\odot\) relative to the SPIPS-implied radius, compared with \(2.060\,R_\odot\) for the larger-radius case. These results show that the observed visual morphology, radial displacement, and temperature-driven luminosity variation are mutually consistent at the few-percent level when placed on a common phase convention, while the absolute radius scale remains the dominant systematic.
\end{abstract}


\section{Introduction}

Classical Cepheids are fundamental calibrators of the extragalactic distance scale, but detailed modeling of individual benchmark Cepheids remains important for understanding the physical origin of their observed light, radial-velocity, radius, and temperature variations. The prototype \(\delta\) Cephei is especially valuable because it has an accurately measured pulsation period, extensive photometry, radial velocities, and published interferometric and spectroscopic constraints.

Cepheids remain central to the calibration of the extragalactic distance scale and the local Hubble constant \citep{freedman2001hstkey,freedman2010hubble,riess2016,riess2022}. The Baade--Wesselink and infrared surface-brightness approaches provide quasi-geometric routes to Cepheid radii and distances by combining photometric, spectroscopic, and angular-diameter information \citep{baade1926,wesselink1946,barnes1976surfacebrightness,fouque1997irsb,storm2011irsb1,storm2011irsb2}. Long-baseline interferometry and related angular-diameter measurements have made nearby Cepheids, especially \(\delta\) Cephei, useful for testing projection factors and radius scales \citep{armstrong2001diameters,kervella2004,merand2005bw,benedict2002deltacep,benedict2007hstcepheids}. Projection-factor systematics remain a central limitation in Baade--Wesselink analyses, motivating both spectroscopic and global modeling studies \citep{nardetto2007pFactor,nardetto2009pFactor,storm2011irsb1,merand2005bw,trahin2021gaiaedr3spips}. The SPIPS framework combines radial velocities, photometry, effective temperatures, angular diameters, reddening, and distance constraints in a single model-based pulsation analysis \citep{merand2015spips,cdsMerand2015,trahin2021gaiaedr3spips}. For \(\delta\) Cephei, binarity and Gaia astrometric systematics also affect interpretation of the parallax and radial-velocity zero point \citep{gaiaedr3,kervella2019companions,nardetto2024deltacep_orbit}.

Nonlinear pulsation and evolutionary calculations can match some global properties of Cepheids while still failing to reproduce detailed visual light-curve amplitude and morphology. The present work therefore takes a complementary route: instead of solving the nonlinear pulsation hydrodynamics, we reconstruct what the observed star empirically requires in terms of \(R(\phi)\), \(T_{\rm eff}(\phi)\), \(L(\phi)\), and \(V(\phi)\). The analysis is diagnostic: it does not derive a new distance, projection factor, reddening, angular diameter, or SPIPS solution.

\section{Data and Literature Benchmarks}

The adopted pulsation period is
\begin{equation}
P_{\rm obs}=5.366531~{\rm d}.
\end{equation}
The empirical photometric analysis uses cleaned Johnson-\(V\) photometry and derives its own Fourier template. The previously used project-level visual amplitude, \(\Delta V_{\rm obs}=0.8390~{\rm mag}\), is retained only as a reference scale. The final template values used in this paper are those measured from the present cleaned data set and listed in Table~\ref{tab:core_results}.

The radial-velocity analysis uses the HARPS-N 2019 radial velocities presented by \citet{nardetto2024deltacep_orbit}. The SPIPS model curves are taken from the public SPIPS/CDS data products for \(\delta\) Cephei \citep{merand2015spips,cdsMerand2015}. In particular, the published SPIPS model table provides phase-dependent radial velocity, pulsation velocity, angular diameter, effective temperature, luminosity, and surface gravity. These curves are used as an external literature benchmark, not as a refitted model.

The absolute radius scale is treated as an adopted input. Two mean-radius cases are tested, \(R_0=43.7\,R_\odot\) and \(R_0=44.9\,R_\odot\). The projection factor is also adopted rather than fitted; the primary radius-displacement calculation quoted below uses \(p=1.317\).

\section{Analysis Method}

\subsection{Phase Folding and Photometric Fourier Model}

For an observation at time \(t\), the pulsation phase is
\begin{equation}
\phi = \left[\frac{t-T_0}{P}\right]\bmod 1,
\end{equation}
where \(T_0\) is the adopted reference epoch and \(P\) is the pulsation period. The Johnson-\(V\) phase curve is represented as
\begin{equation}
V(\phi)=A_0+\sum_{k=1}^{N} A_k\cos(2\pi k\phi+\psi_k),
\end{equation}
with Fourier amplitude ratios
\begin{equation}
R_{k1}=\frac{A_k}{A_1}.
\end{equation}
The adopted empirical template uses \(N=3\) harmonics after median observer-offset correction and residual clipping.

\subsection{Radial Velocity, Radius Displacement, and Phase Alignment}

The pulsational radial-velocity curve was represented by a smooth Fourier fit to the HARPS-N velocities. For the radius-displacement diagnostic, the fitted mean velocity is subtracted before integration. The photospheric radius variation is then written as
\begin{equation}
\Delta R(\phi)=
-pP\int_{\phi_0}^{\phi} v_{\rm rel}(\phi')\,d\phi',
\end{equation}
where \(v_{\rm rel}\) is the mean-subtracted velocity curve and \(p\) is the adopted projection factor. The full radius curve is
\begin{equation}
R(\phi)=R_0+\Delta R(\phi).
\end{equation}
Because the HARPS-N and SPIPS phase conventions are not identical, the HARPS-N radial-velocity curve was aligned to the SPIPS radial-velocity curve by fitting a phase shift and velocity offset. This alignment is used only to sample SPIPS \(T_{\rm eff}(\phi)\), \(L(\phi)\), and related quantities on the HARPS-N radius-curve phase grid.

\subsection{Hybrid Luminosity and SPIPS-Implied Radius}

Given a phase-dependent radius and effective temperature, the luminosity is
\begin{equation}
\frac{L(\phi)}{L_\odot}
=
\left[\frac{R(\phi)}{R_\odot}\right]^2
\left[\frac{T_{\rm eff}(\phi)}{T_\odot}\right]^4.
\end{equation}
For the hybrid test, \(R(\phi)\) is the preliminary HARPS-N-based radius curve and \(T_{\rm eff}(\phi)\) is the published SPIPS effective-temperature curve. The radius implied directly by the published SPIPS luminosity and temperature curves is
\begin{equation}
R_{\rm SPIPS}(\phi)
=
\frac{\sqrt{L_{\rm SPIPS}(\phi)/L_\odot}}
{\left[T_{{\rm eff,SPIPS}}(\phi)/T_\odot\right]^2}
\,R_\odot .
\end{equation}
This quantity provides an independent check on which adopted mean-radius scale is more consistent with the SPIPS luminosity and temperature scale.

\section{Empirical Johnson-\texorpdfstring{\(V\)}{V} Light-Curve Template}

The cleaned AAVSO Johnson-\(V\) photometry was phase-folded using the adopted pulsation period \(P_{\rm obs}=5.366531~{\rm d}\). A median observer-offset correction and iterative residual clipping were applied before fitting a three-harmonic Fourier template. The resulting empirical template has \(\Delta V=0.832939~{\rm mag}\), \(R_{21}=0.382395\), \(R_{31}=0.167631\), and rise fraction \(0.286800\). This empirical template provides the observed visual morphology against which the SPIPS benchmark and the semi-empirical reconstruction are compared.

\begin{figure}[htbp]
\centering
\includegraphics[width=0.88\textwidth]{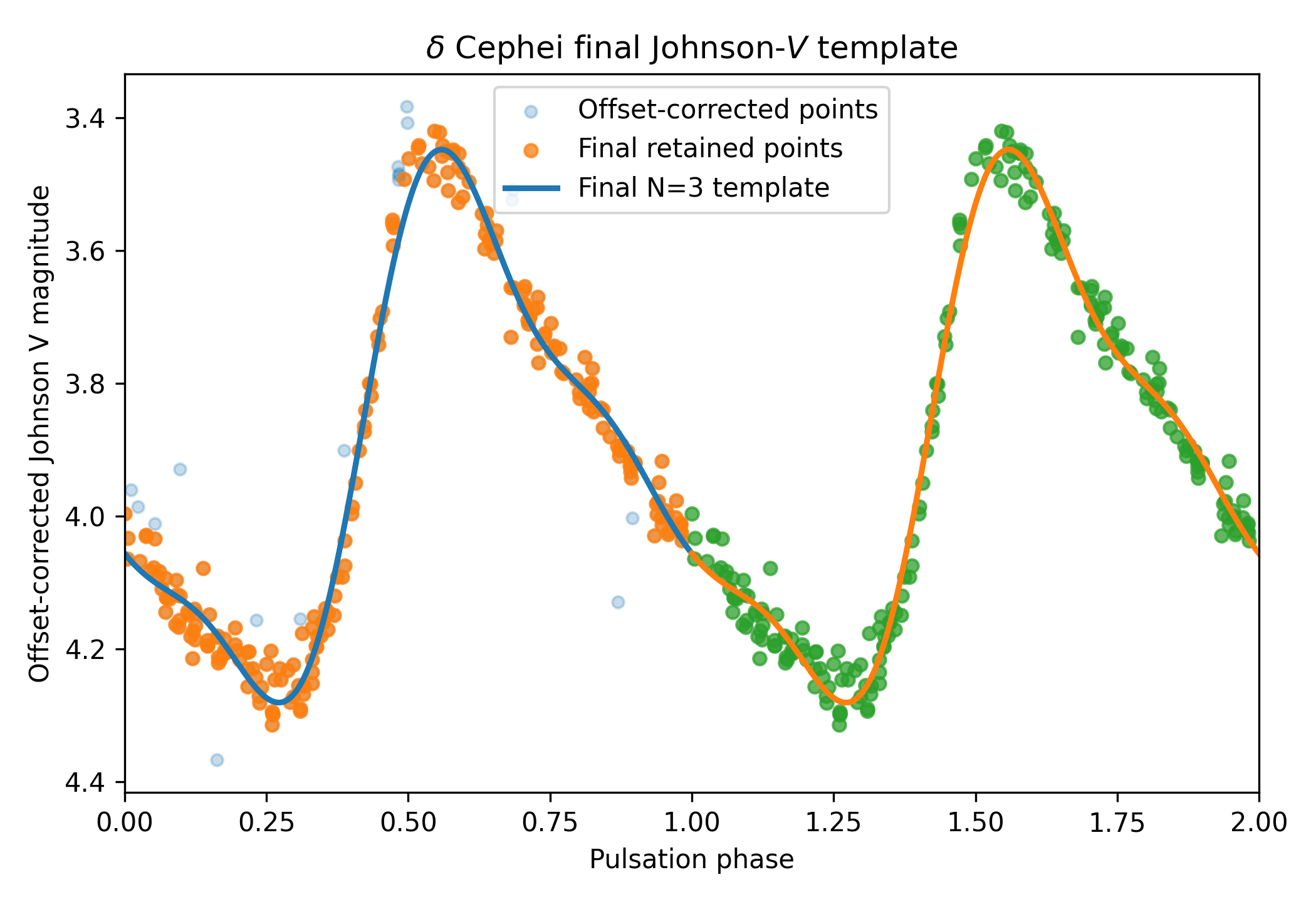}
\caption{Empirical Johnson-\(V\) light-curve template for \(\delta\) Cephei after observer-offset correction and residual clipping. The solid curve is the adopted three-harmonic Fourier template.}
\label{fig:johnson_v_template}
\end{figure}

\section{SPIPS Benchmark Comparison}

The public SPIPS/CDS model table for \(\delta\) Cephei was used as an external literature benchmark, not as a new fit \citep{merand2015spips,cdsMerand2015}. The SPIPS \(V\)-band curve was compared with the empirical Johnson-\(V\) template after allowing only a phase shift and a vertical magnitude offset. After this alignment, the SPIPS visual curve reproduces the empirical morphology with an RMS residual of \(0.023456~{\rm mag}\), a median absolute residual of \(0.017099~{\rm mag}\), and a maximum absolute residual of \(0.055821~{\rm mag}\). The SPIPS raw \(V\)-band amplitude is \(0.873850~{\rm mag}\), which is \(0.040911~{\rm mag}\) larger than the empirical template amplitude.

\begin{figure}[htbp]
\centering
\includegraphics[width=0.88\textwidth]{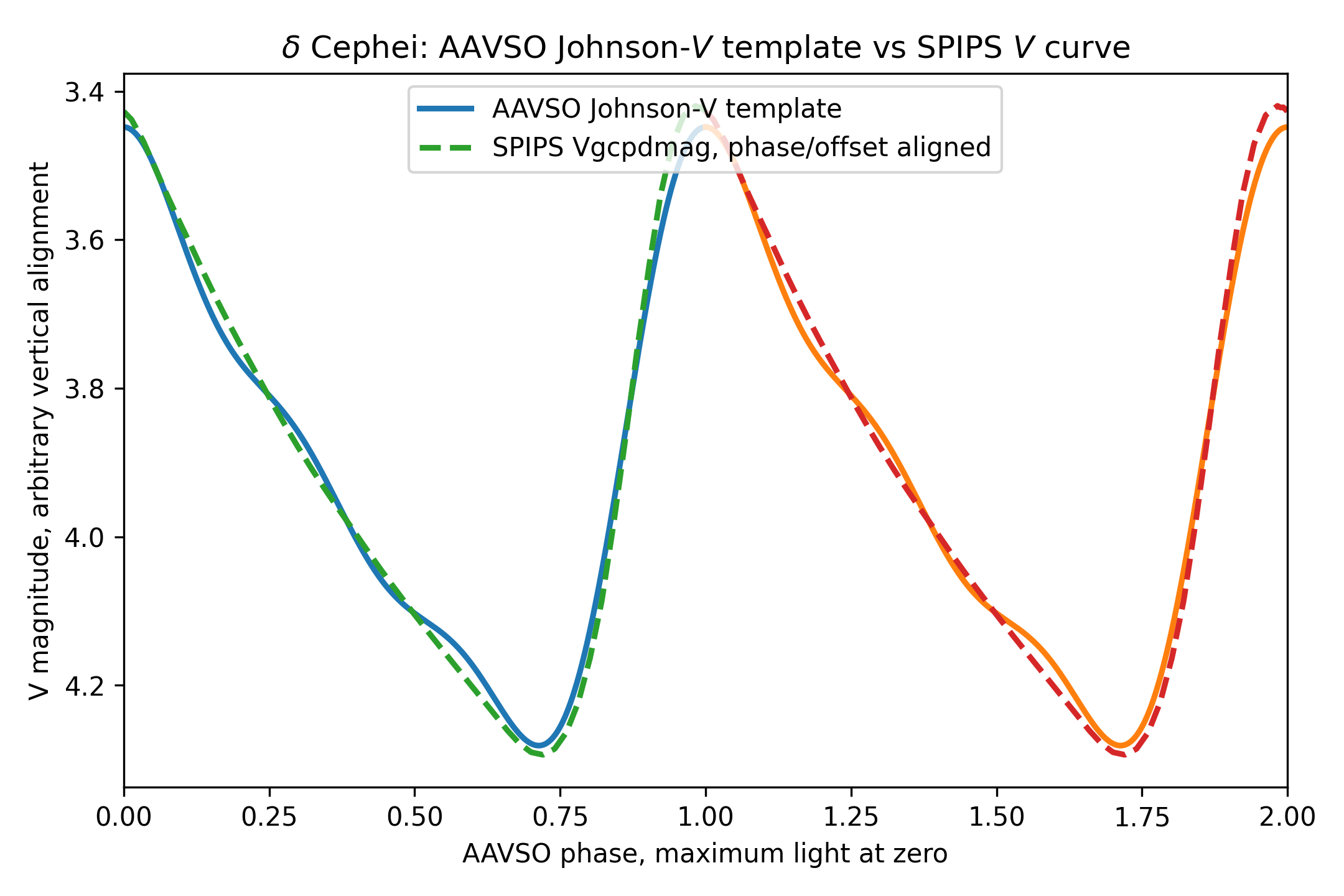}
\caption{Comparison between the empirical Johnson-\(V\) template and the published SPIPS \(V\)-band curve after phase and vertical alignment. The small residuals show that the published SPIPS model reproduces the observed visual morphology well without refitting the model.}
\label{fig:spips_vs_v_template}
\end{figure}

The HARPS-N radial-velocity curve was also compared with the published SPIPS radial-velocity curve to determine the relative phase convention \citep{nardetto2024deltacep_orbit,merand2015spips}. The best phase shift added to the HARPS-N phase to sample the SPIPS curve is \(0.871667\) cycle, equivalent to \(-0.128333\) cycle modulo one period. The shape RMS residual between the aligned velocity curves is \(1.052357~{\rm km\,s^{-1}}\), indicating that the two curves have closely consistent phase dependence once the zero-point convention is removed.

\begin{figure}[htbp]
\centering
\includegraphics[width=0.88\textwidth]{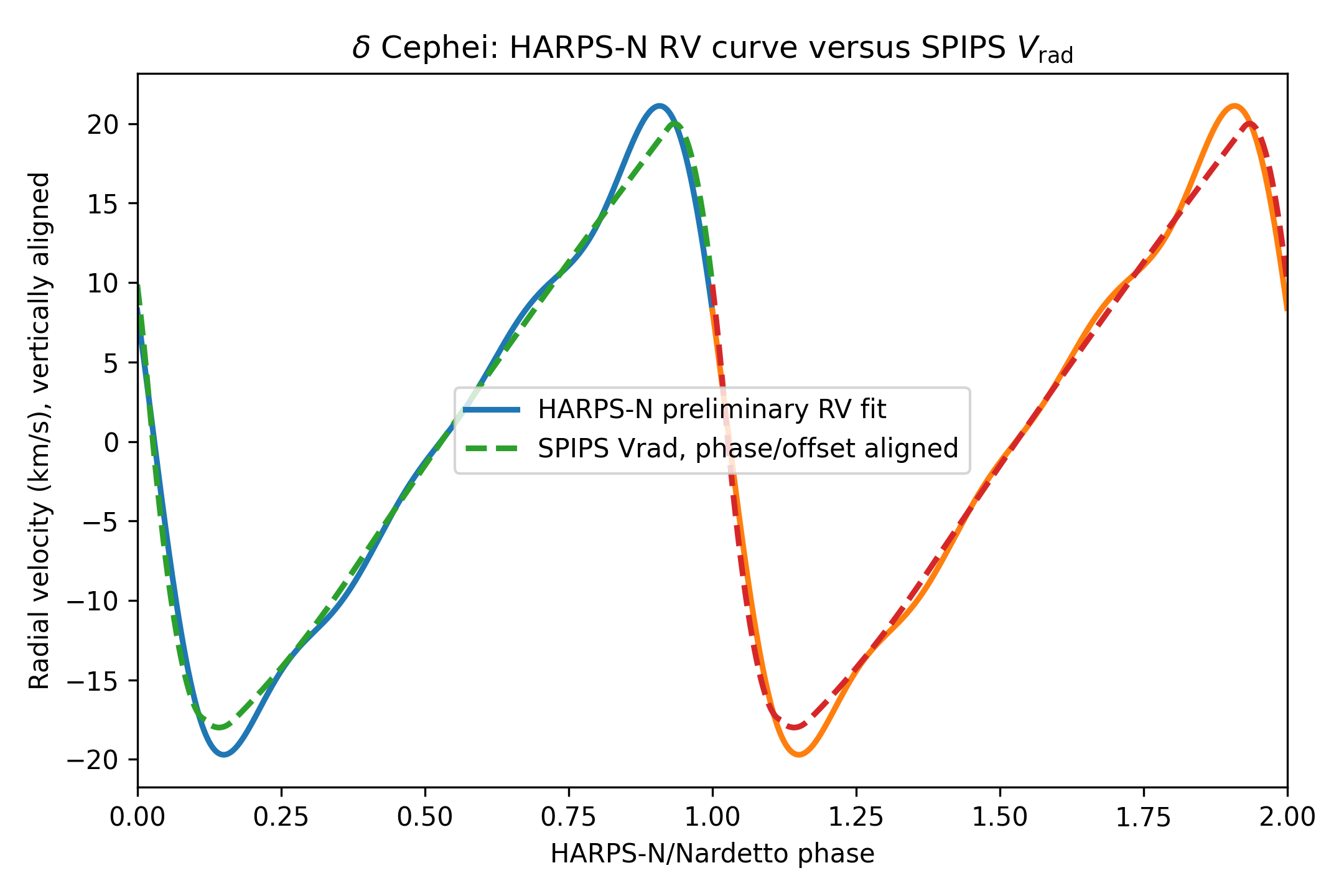}
\caption{Phase alignment between the HARPS-N radial-velocity curve and the published SPIPS radial-velocity curve. The comparison is used to place the preliminary radius-displacement curve and SPIPS temperature curve on a common phase grid.}
\label{fig:rv_phase_alignment}
\end{figure}

\section{Radius Displacement and Hybrid Luminosity Reconstruction}

The HARPS-N radial velocities were fitted with a smooth Fourier representation. The fitted peak-to-peak velocity span is \(40.819572~{\rm km\,s^{-1}}\). Integrating the mean-subtracted curve with \(p=1.317\) gives a preliminary radius-displacement peak-to-peak amplitude of \(4.809677\,R_\odot\). This displacement amplitude is robust to the tested velocity zero-point and orbital-correction conventions because the fitted mean velocity is subtracted before integration. The calculation follows the standard Baade--Wesselink logic \citep{baade1926,wesselink1946,merand2005bw,nardetto2009pFactor,storm2011irsb1}, but the absolute radius scale remains an adopted input.

The preliminary HARPS-N-based radius curve was then combined with the published SPIPS \(T_{\rm eff}(\phi)\) curve to compute a hybrid luminosity curve. The SPIPS temperature sampled on the aligned HARPS-N phase grid ranges from \(5311.725\) to \(6351.110~{\rm K}\), while the published SPIPS luminosity sampled on the same grid ranges from \(1290.876\) to \(2458.472\,L_\odot\). For the \(R_0=43.7\,R_\odot\) case, the hybrid luminosity ranges from \(1341.605\) to \(2540.194\,L_\odot\) and has mean ratio \(L_{\rm hybrid}/L_{\rm SPIPS}=1.040269\). For the \(R_0=44.9\,R_\odot\) case, the corresponding mean ratio is \(1.098261\). Thus the smaller adopted radius scale is more consistent with the published SPIPS luminosity scale.

\begin{figure}[htbp]
\centering
\includegraphics[width=0.88\textwidth]{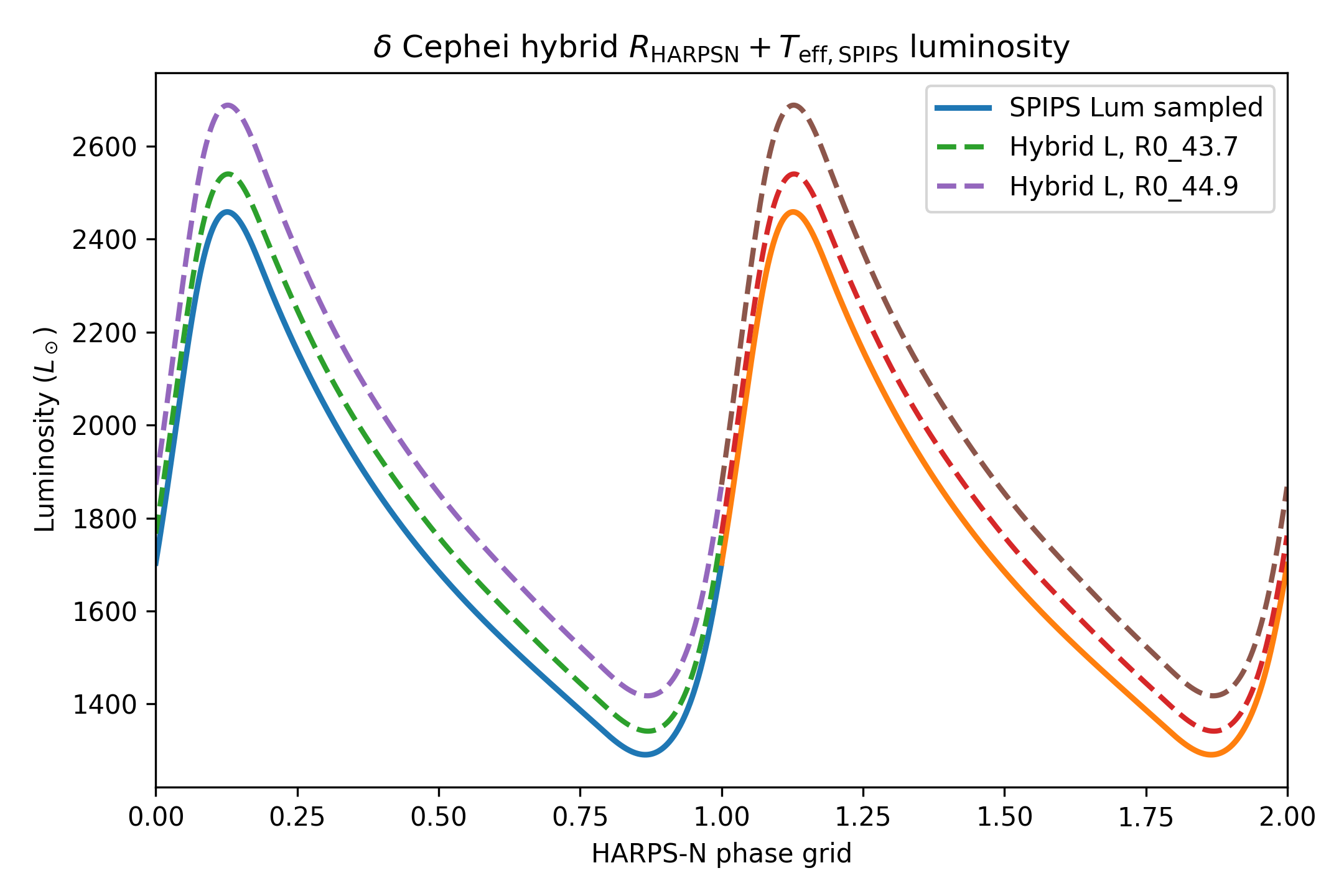}
\caption{Hybrid luminosity curves obtained from the preliminary HARPS-N radius curves and the published SPIPS effective-temperature curve, compared with the published SPIPS luminosity curve.}
\label{fig:hybrid_luminosity}
\end{figure}

\section{Radius-Scale Comparison}

The SPIPS-implied radius curve computed from \(L_{\rm SPIPS}(\phi)\) and \(T_{{\rm eff,SPIPS}}(\phi)\) ranges from \(40.043\) to \(44.615\,R_\odot\), with mean radius \(42.843\,R_\odot\) and peak-to-peak amplitude \(4.573\,R_\odot\). The HARPS-N radius curve with \(R_0=43.7\,R_\odot\) ranges from \(40.784\) to \(45.594\,R_\odot\), giving a mean offset of \(+0.857\,R_\odot\) and RMS offset of \(0.864\,R_\odot\) relative to the SPIPS-implied radius. The \(R_0=44.9\,R_\odot\) case ranges from \(41.984\) to \(46.794\,R_\odot\), giving a larger mean offset of \(+2.057\,R_\odot\) and RMS offset of \(2.060\,R_\odot\). The comparison therefore favors the lower adopted radius scale.

\begin{figure}[htbp]
\centering
\includegraphics[width=0.88\textwidth]{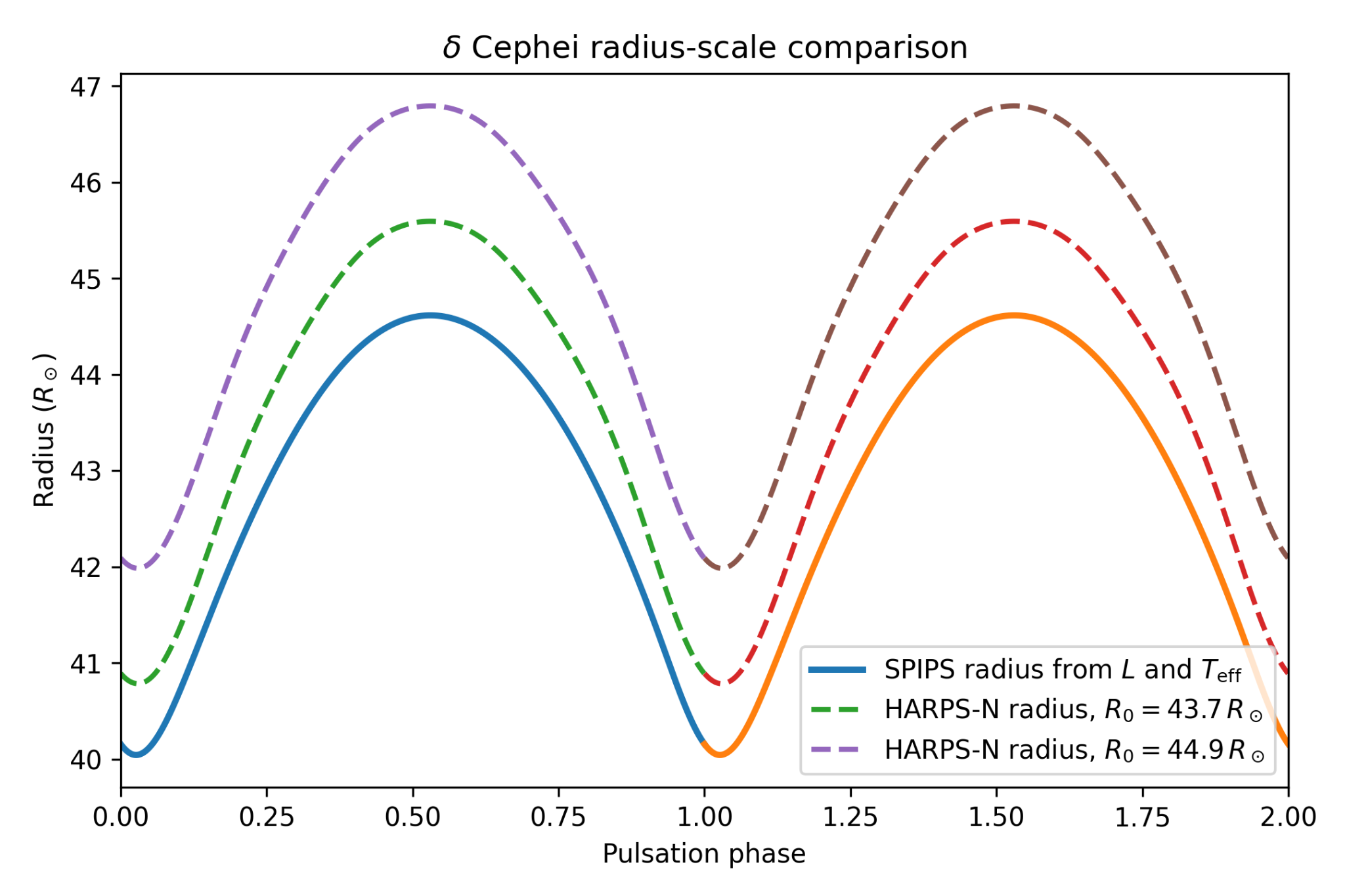}
\caption{Radius-scale comparison between the SPIPS-implied radius curve and preliminary HARPS-N radius curves with two adopted mean-radius scales. The \(R_0=43.7\,R_\odot\) case is closer to the SPIPS-implied radius scale than the \(R_0=44.9\,R_\odot\) case.}
\label{fig:radius_scale_comparison}
\end{figure}

\begin{table}[htbp]
\centering
\caption{Core semi-empirical reconstruction results.}
\label{tab:core_results}
\small
\begin{tabular}{llc l}
\toprule
Category & Quantity & Value & Unit \\
\midrule
Observed \(V\) template & \(\Delta V\) & 0.832939 & mag \\
Observed \(V\) template & \(R_{21}\) & 0.382395 &  \\
Observed \(V\) template & \(R_{31}\) & 0.167631 &  \\
Observed \(V\) template & rise fraction & 0.286800 & cycle \\
SPIPS \(V\) comparison & shape RMS residual & 0.023456 & mag \\
SPIPS \(V\) comparison & \(\Delta V_{\rm SPIPS}-\Delta V_{\rm emp}\) & 0.040911 & mag \\
HARPS-N RV & RV peak-to-peak & 40.819572 & km\,s\(^{-1}\) \\
Radius variation & \(\Delta R\) peak-to-peak, \(p=1.317\) & 4.809677 & \(R_\odot\) \\
Hybrid luminosity & mean \(L_{\rm hybrid}/L_{\rm SPIPS}\), \(R_0=43.7\,R_\odot\) & 1.040269 &  \\
Hybrid luminosity & mean \(L_{\rm hybrid}/L_{\rm SPIPS}\), \(R_0=44.9\,R_\odot\) & 1.098261 &  \\
Radius scale & RMS \(R-R_{\rm SPIPS}\), \(R_0=43.7\,R_\odot\) & 0.863737 & \(R_\odot\) \\
Radius scale & RMS \(R-R_{\rm SPIPS}\), \(R_0=44.9\,R_\odot\) & 2.059669 & \(R_\odot\) \\
\bottomrule
\end{tabular}
\end{table}

\section{Discussion}

The empirical Johnson-\(V\) template, the published SPIPS visual curve, the HARPS-N radial velocities, and the SPIPS temperature and luminosity curves give a mutually consistent description of the pulsation when placed on a common phase convention. The SPIPS \(V\)-band morphology matches the empirical Johnson-\(V\) template to within a few hundredths of a magnitude after only phase and vertical alignment. This agreement shows that the observed visual morphology is already well represented by the published SPIPS solution and that the current analysis is best interpreted as an independent consistency check rather than a refit.

The radius-displacement calculation provides a complementary constraint. The HARPS-N velocities imply a preliminary displacement amplitude of about \(4.81\,R_\odot\) for \(p=1.317\). When this radius variation is combined with the SPIPS effective-temperature curve, the resulting hybrid luminosity curve closely follows the published SPIPS luminosity curve. The agreement is better for \(R_0=43.7\,R_\odot\) than for \(R_0=44.9\,R_\odot\), both in the direct luminosity comparison and in the SPIPS-implied radius comparison.

The dominant remaining systematic is the absolute radius scale. The radius-displacement amplitude is determined by the radial-velocity integration and the adopted projection factor, whereas the vertical placement of the radius curve depends on the adopted \(R_0\). The present analysis therefore supports the phase-dependent shape and amplitude of the radius variation more strongly than it determines an independent absolute radius. Additional angular-diameter, reddening, distance, and projection-factor constraints would be required for a new absolute Baade--Wesselink solution.

The semi-empirical reconstruction also provides a useful benchmark for nonlinear pulsation calculations. Models that reproduce the pulsation period but fail to match the observed Johnson-\(V\) amplitude or Fourier morphology can now be compared against separate empirical constraints on \(V(\phi)\), \(R(\phi)\), \(T_{\rm eff}(\phi)\), and \(L(\phi)\). This separation helps identify whether discrepancies arise primarily from the radius variation, the temperature variation, atmosphere/bolometric-correction effects, or the adopted absolute scale.

\section{Conclusions}

We constructed a semi-empirical reconstruction of \(\delta\) Cephei using cleaned Johnson-\(V\) photometry, HARPS-N radial velocities, and published SPIPS model curves. The main conclusions are as follows.

\begin{enumerate}
\item The empirical Johnson-\(V\) template has \(\Delta V=0.832939~{\rm mag}\), \(R_{21}=0.382395\), \(R_{31}=0.167631\), and rise fraction \(0.286800\).
\item After phase and vertical alignment, the published SPIPS \(V\)-band curve reproduces the empirical Johnson-\(V\) morphology with RMS residual \(0.023456~{\rm mag}\).
\item The HARPS-N radial velocities give a fitted peak-to-peak velocity span of \(40.819572~{\rm km\,s^{-1}}\) and a preliminary radius-displacement amplitude of \(4.809677\,R_\odot\) for \(p=1.317\).
\item Combining the preliminary HARPS-N radius curve with the SPIPS effective-temperature curve yields a hybrid luminosity curve close to the published SPIPS luminosity curve.
\item The \(R_0=43.7\,R_\odot\) radius scale is more consistent with the SPIPS luminosity and temperature scale than the \(R_0=44.9\,R_\odot\) case, with RMS radius offset \(0.864\,R_\odot\) instead of \(2.060\,R_\odot\).
\end{enumerate}

These results show that the observed visual morphology, radial displacement, and temperature-driven luminosity variation of \(\delta\) Cephei are consistent at the few-percent level when placed on a common phase convention. The analysis does not constitute a new SPIPS fit or a new absolute Baade--Wesselink radius measurement; rather, it provides an observationally anchored benchmark for interpreting Cepheid pulsation models.

\section*{Acknowledgments}

This work made use of observations from the AAVSO International Database contributed by observers worldwide. The authors thank the AAVSO observers whose Johnson-\(V\) measurements made the empirical light-curve template possible. This work also made use of published SPIPS model data and HARPS-N radial velocities from the literature.

\section*{Data Availability}

The AAVSO photometry used in this work is available through the AAVSO International Database. The SPIPS model curves and HARPS-N radial velocities are available from the cited literature and associated public data products. Processed tables and scripts can be provided upon reasonable request.

\bibliographystyle{unsrtnat}
\bibliography{paper9}

\end{document}